\documentclass[twocolumn]{aastex631}

\usepackage{newtxtext,newtxmath}

\usepackage[T1]{fontenc}

\DeclareRobustCommand{\VAN}[3]{#2}
\let\VANthebibliography\thebibliography
\def\thebibliography{\DeclareRobustCommand{\VAN}[3]{##3}\VANthebibliography}

\usepackage{graphicx}	
\usepackage{amsmath}	
\usepackage{amssymb}	
\usepackage{color}
\usepackage{xcolor}

\newcommand{\LCDM}{\rm{\Lambda}CDM}

\begin{document}

\title{Redshift evolution of the X-ray and UV luminosity relation of quasars: calibrated results from SNe Ia}

\author{Xiaolei Li}
\affiliation{College of Physics, Hebei Normal University, Shijiazhuang 050024, China}

\author{Ryan E. Keeley}
\affiliation{Department of Physics, University of California Merced, 5200 North Lake Road, Merced, CA 95343, USA}

\author{Arman Shafieloo}
\affiliation{Korea Astronomy and Space Science Institute, Daejeon 34055, Korea}
\affiliation{University of Science and Technology, Yuseong-gu 217 Gajeong-ro, Daejeon 34113, Korea}




\begin{abstract}
Quasars could serve as standard candles if the relation between their ultraviolet and X-ray luminosities can be accurately calibrated. Previously, we developed a model-independent method to calibrate quasar standard candles using the distances-redshift relation reconstructed from Type Ia supernova at $z<2$ using Gaussian process regression.
Interestingly, we found that the calibrated quasar standard candle dataset preferred a deviation from $\LCDM$ at redshifts above $z>2$. One possible interpretation of these findings is that the calibration parameters of the quasar ultraviolet and X-ray luminosity relationship evolves with redshift. In order to test the redshift dependence of the quasar calibration in a model-independent manner, we divided the quasar sample whose redshift overlap with the redshift coverage of Pantheon+ Type Ia supernova compilation into two sub-samples: a low-redshift quasar sub-sample and a high-redshift quasar sub-sample. {Assuming all the quasar samples are reliable,} our results show that there is about a 4$\sigma$ inconsistency between the quasar parameters inferred from the sub-samples without considering evolution. This inconsistency suggests the possibility of considering redshift evolution for the relationship between the quasars’ ultraviolet and X-ray luminosities. We then test an explicit parametrization of the redshift evolution of the quasar calibration parameters via $\gamma(z)=\gamma_0+\gamma_1(1+z)$ and $\beta(z)=\beta_0+\beta_1(1+z)$. Combining this redshift-dependent calibration relationship with the distance-redshift relationship reconstructed from Pantheon+ supernova compilation, we find the high-redshift sub-sample and low-redshift sub-sample become consistent at the 2$\sigma$ level, which means that the parameterized form of $\gamma(z)$ and $\beta(z)$ works well at describing the evolution of the quasar calibration parameters.

\end{abstract}




\section{Introduction}\label{sec:intro}
Because quasars are among the most luminous persistent sources in the Universe, they can be detected out to redshifts $z\sim 7$~\citep{Mortlock:2011va,Banados:2017unc,2021ApJ...907L...1W}. 
Thus if quasars can be used as standard candles, then they can measure the expansion rate of the Universe beyond the redshifts probed by Type Ia supernovae (SNe Ia) and help us to better understand dark energy. The standard cosmological scenario, the so-called $\Lambda$-cold dark matter ($\LCDM$) model, accurately describes the physics of the early Universe, the formation of large-scale structure, as well as the late-time acceleration of the Universe.
Though this model suffers from some profound theoretical difficulties, including the fine-tuning problem and the coincidence problem, as well as observational tensions including the Hubble tension and S8 tension \citep{Dolgov:1997za,Weinberg:1988cp,Carroll:1991mt,Martin:2012bt,Aghanim:2018eyx,Riess_2019,DiValentino:2021izs}.

Historically, there have been a number of attempts to use quasars as standard candles from a variety of emission features.
For example, \cite{1977ApJ...214..679B} showed that the width of quasars' emission lines were correlated with their luminosity, thus demonstrating an avenue by which quasars could be standard candles, however in \cite{1999ASPC..162..235O}, the authors argued that the ``substantial scatter'' in this relationship makes it not useful for cosmology.
Furthermore, in \cite{Wang:2013ha}, the authors demonstrated that super-Eddington accreting massive black holes have a luminosity determined by the mass of the black hole and thus that these objects are standard candles. \cite{2014ApJ...787L..12L} showed that the X-ray variability of quasars can be used to measure their luminosities and in the work of \cite{Watson:2011um, Melia:2013sxa, 2015ApJ...801....8K}, the authors show that the radius of a quasar can be used to measure its luminosity.

Despite these multiple avenues by which quasars could serve as standard candles, the observed relationship between the log of the X-ray and ultraviolet (UV) luminosities~\citep{1979ApJ...234L...9T,1981ApJ...245..357Z,1986ApJ...305...83A,2010A&A...512A..34L,2010A&A...519A..17V} has recently be widely used to measure cosmological distances~\citep{Risaliti:2015zla,Risaliti:2016nqt,Lusso:2017hgz,Risaliti:2018reu,Lusso:2020obu,Khadka:2020vlh,Khadka:2020tlm}.

Initially, this dataset of quasar fluxes was found to fit the $\LCDM$ model well and matched the distances inferred from SNe Ia~\citep{Risaliti:2015zla}. 
However, various groups found that there exists a deviation between the $\LCDM$ model and the quasar distances when they are calibrated with such a linear relationship~\citep{Risaliti:2018reu,Khadka:2020tlm,Li:2021onq}.
For instance, \citet{Risaliti:2018reu} found that the best-fit value of the present matter density ($\Omega_{{\rm{m}},0}$) is different if constrained with only high-redshift quasars or only low-redshift quasars.
Furthermore, \citet{Khadka:2020tlm} found that the inferred quasar calibration parameters depend on which cosmological is used, concluding that only quasars in the range $z<1.5-1.7$ are reliable probes of cosmological distances.
Also, \citet{Li:2021onq} found that a model-independent calibration of quasars can be performed out to $z<2$, that calibrated quasar datset deviated from $\LCDM$ at $z>2$.
In \cite{Sacchi:2022ofz}, the authors performed a one-by-one analysis of a sample of 130 quasars at $z > 2.5$ with high-quality X-ray and UV spectroscopic observations. The results show that the X-ray-UV calibration still holds at $z > 2.5$. On the other hand, in the work of \cite{Li:2022inq}, the authors combine the PAge approximation (Parameterization based on cosmic Age) and a high-quality quasar sample to search for the origins of the deviation and found that the deviation from the standard $\LCDM$ model probably originates from the redshift-evolution effects and non-universal intrinsic dispersion of the quasar luminosity relation rather than new cosmological physics. Moreover, in \cite{Wang:2022hko}, the authors constructed a three-dimensional and redshift-evolutionary X-ray and UV luminosity relation for quasars from the powerful statistic tool called copula, and found that the constructed $L_{\rm{X}}-L_{\rm{UV}}$ relation from copula is more viable and the observations favored their modified X-ray and UV luminosity relation with the additional redshift dependent term. Their results confirmed that the quasars can be regarded as a reliable indicator of the cosmic distance if the $L_{\rm{X}}-L_{\rm{UV}}$ relation
from copula is used to calibrate quasar data.
In a recent work by \cite{wang2024observations}, they compare three different quasar $L_X-L_{UV}$ relations (two $L_X-L_{UV}$ relation includes no redshift-evolutionary while two $L_X-L_{UV}$ relations allow for redshift evolution) and conclude that the $L_X-L_{UV}$ relations that allow for redshift evolution are favored by the observations. 
By dividing the quasar sample into a low-redshift sub-sample and a high-redshift sub-sample, the constrain results on PAge parameters are consistent for the redshift-evolutionary relation while the constrain results about quasar parameters, i.e., $\gamma$ and $\beta$ are still in tensions.

These works collectively indicate that there is some tension between the quasar data and the joint model including both the standard UV and X-ray luminosity relationship and the $\LCDM$ model.
This tension could either be explained by extending the cosmological model beyond $\LCDM$, or by adding a redshift evolution to the quasar calibration parameters.

Different methods have been used in the literature to calibrate quasar distances. {For instance, \cite{Risaliti:2015zla} use narrow redshift bins of quasars to estimate $\gamma$ in that narrow redshift bin and then use the average value of $\gamma$ over the redshift bins to calculate cosmological constraints.
Also, \cite{Lusso:2020pdb} fit the luminosity distance with a fifth-grade polynomial in $\log (1+z)$ in order to calibrate quasar UV and X-ray luminosity relation.}
In this work, we follow the method developed in \cite{Li:2021onq} where the quasar UV and X-ray luminosity relationship is calibrated in a cosmological-model-independent manner. This is achieved by using Gaussian process (GP) regression to reconstruct the distance-redshift relationship from SNe Ia and then varying the quasar calibration parameters such that these reconstructions are maximally consistent with the quasar data. We use this methodology to test if different high-redshift and low-redshift sub-samples of the quasar data are consistent with each other. We also use this methodology to test a new quasar calibration parametrization that evolves with redshift, thus accounting for the tension between high-redshift and low-redshift subsets of the quasar dataset.

\section{Calibration Method and quasar sample} \label{sec:cali}
The standard relationship between quasars' X-ray and UV luminosities is parameterized as 
\begin{equation}\label{eq:logL}
     {\rm{log }}(L_{\rm{X}})\,=\,\gamma {\rm{log }}(L_{\rm{UV}})+\beta,
\end{equation}
where $\gamma$ and $\beta$ are the two quasar calibration parameters that are fit to the data, and $\log \equiv \log_{10}$.
With the relation between the luminosity $L$ and the flux $F$: $L = 4\pi D_{\rm{L}}^2 F$, 
we can rewrite equation~(\ref{eq:logL}) as:
\begin{equation} \label{eq:logFx1}
    {\log (F_{\rm{X}})}\,=\,\gamma \log (F_{\rm{UV}}) + (2\gamma-2){\rm{log }}(D_{\rm{L}}) + (\gamma-1) \log (4\pi)+\beta,
\end{equation}
where $F_{\rm{UV}}$ and $F_{\rm{X}}$ are the fluxes measured at fixed rest-frame wavelengths, and $D_{\rm{L}}$ is the luminosity distance. From this relation, we can see that the quasar calibration parameters are degenerate with the cosmological distances. Thus, any observed deviation in the data could potentially be explained by either extending the cosmological model, or by extending the quasar calibration relation. In keeping with the theme of developing statistical methods to make joint cosmological constraints from a variety of datasets, we use the distances from SNe Ia and investigate whether an extension in the quasar calibration relation is needed by the data. This is, in essence, why we find extending the calibration model more convincing than extending the cosmological model; extending the cosmological model to explain this tension in the quasar data would be inconsistent with the results from SNe Ia. This also highlights the importance of model-independent methods; we are as flexible and as agnostic, with regards to the true cosmology, as the data allow.

Following the work of \cite{Li:2021onq}, we use cosmological distances from SNe Ia to break the degeneracy between the quasar calibration parameters and cosmological distances and thus constrain the quasar calibration parameters. We generate 1000 realizations of the unanchored luminosity distance $D_{\rm{L}}H_0$ from the posterior of the Pantheon+ compilation from \cite{Scolnic:2021amr} calculated with GP regression. {
It is worth noting that the GP reconstruction is weakly dependent on the chosen mean function, especially in regions that the data do not constrain well. Different mean functions will cause the data to prefer different regions of the hyperparameter space which can effect the uncertainty in the reconstruction where the data do not constrain it well.
For comprehensive details on this sampling method, we refer the readers to \cite{KeeleySLa, KeeleySLb}, and for an extensive discourse on GP regression, please refer to \cite{Rasmussen:2006,Holsclaw:2010nb,2011PhRvD..84h3501H, Shafieloo2012Gaussian,KeeleyGP,Hwang:2022hla}.} We then calculate the predicted quasar X-ray flux ($F_{\rm{X}}^{\rm{pre}}$) corresponding to these unanchored luminosity distances $D_{\rm{L}}H_0$, along with the parameters of the standard quasar calibration relation, by rewriting equation~(\ref{eq:logFx1}) as,
\begin{equation} \label{eq:logFx}
    \begin{aligned}
      \log (F_{\rm{X}}^{\rm{pre}})\,=\, & \gamma \log (F_{\rm{UV}}) + (2\gamma-2){\rm{log }}(D_{\rm{L}}H_0)\\
      &-(2\gamma-2)\log (H_0)\\  
      & + (\gamma-1) \log (4\pi)+\beta. 
    \end{aligned}
\end{equation}
This allows us to compare the observed values of the quasars' X-ray flux ${\rm{log }}(F_{\rm{X}})$ and the values predicted by the quasar calibration parameters and cosmological parameters.

Then, following \cite{Risaliti:2015zla,Lusso:2020pdb}, we define the likelihood ($\mathcal{L} = \exp\left( -\chi^2 /2 \right)$) of the quasar parameters based on a modified $\chi^2$ function, which includes a penalty term for the intrinsic dispersion $\delta$, 
\begin{equation}\label{eq:chi2}
    \chi^2\,=\,\sum_i \left[ \frac{\left(\log (F_{\rm{X}}(\gamma,\beta,H_0))^{\rm{pre}}_i- \log (F_{\rm{X}})_i^{\rm{obs}}\right)^2} {s_i^2}  +{\rm{ln}} (s_i^2)\right],
\end{equation}
where $s_i^2\,=\,\sigma_{\log (F_{\rm{X}})}^2+\gamma^2 \sigma_{\log (F_{\rm{UV}})}^2+\delta^2$. 
The intrinsic dispersion $\delta$ is considered in order to allow for scatter in the $L_{\rm{X}}-L_{\rm{UV}}$ relation beyond any measurement noise~\citep{Risaliti:2018reu,Lusso:2020pdb}.

When calibrating the standard quasar calibration relation, $L_{\rm{X}}-L_{\rm{UV}}$, there are a total of four free parameters that need to be constrained, i.e., the slope $\gamma$, the intercept $\beta$ and the intrinsic dispersion parameter $\delta$, 
as well as the Hubble constant $H_0$, 
as seen in equation~({\ref{eq:logFx}}) and equation~(\ref{eq:chi2}).

In addition, we also consider the case where the quasar calibration parameters, $\gamma$ and $\beta$, can evolve with redshift as {$\gamma (z)\,=\, \gamma_0+\gamma_1\times (1+z)$ and $\beta (z)\,=\, \beta_0+\beta_1\times (1+z)$}, respectively. By considering the redshift evolution of the quasar calibration parameters, we now have six parameters, i.e., $\gamma_0$, $\gamma_1$, $\beta_0$, $\beta_1$, $\delta$ and $H_0$, to be constrained.

The statistical analysis is performed by using the Markov Chains Monte Carlo (MCMC) method as implemented in the Python package \texttt{emcee}~\citep{foreman2013emcee} and the converged chains are analyzed with \texttt{GetDist} \citep{Lewis:2019xzd}.

During our analysis, we use the 2421 quasar sample compiled in \citep{Lusso:2020pdb} with redshifts spanning $0.009<z<7.6$. In order to check the redshift dependence of the quasar parameters, we divided the quasar sample into two sub-samples and do the calibration following the method described above. There are 2067 quasars in the sample whose redshift lies within the redshift range of the SNe Ia and thus can be used to do the calibration directly. 
We first divided the 2067 quasars sample into two similar groups: a lower half which has 1033 quasars with $z<1.16$ (hereafter low-redshift sub-sample) and a higher half which has 1034 quasars with $z>1.16$ (hereafter high-redshift sub-sample). Furthermore, we also calibrate the quasar parameters using all the 2067 quasars that have redshifts in the range covered by SNe Ia.

{We should emphasize here that, quasars at redshift lower than $0.7$ may be less reliable due to some residual contamination from the host galaxy~\citep{Lusso:2020pdb}. One can also add a filter of $z>0.7$ to the full sample. In our analysis, we use the full sample to check the redshift dependence for the UV and X-ray luminosity relationship. However, we show the calibrated results for the sample discarding quasars at $z<0.7$ in the appendix. Importantly, the high-redshift subsample prefers an evolution in the quasar calibration parameters regardless of any $z<0.7$ contamination.}

\section{Calibration Results and Discussions}\label{sec:res}

\begin{figure}
\centering
\includegraphics[width=0.47\textwidth]{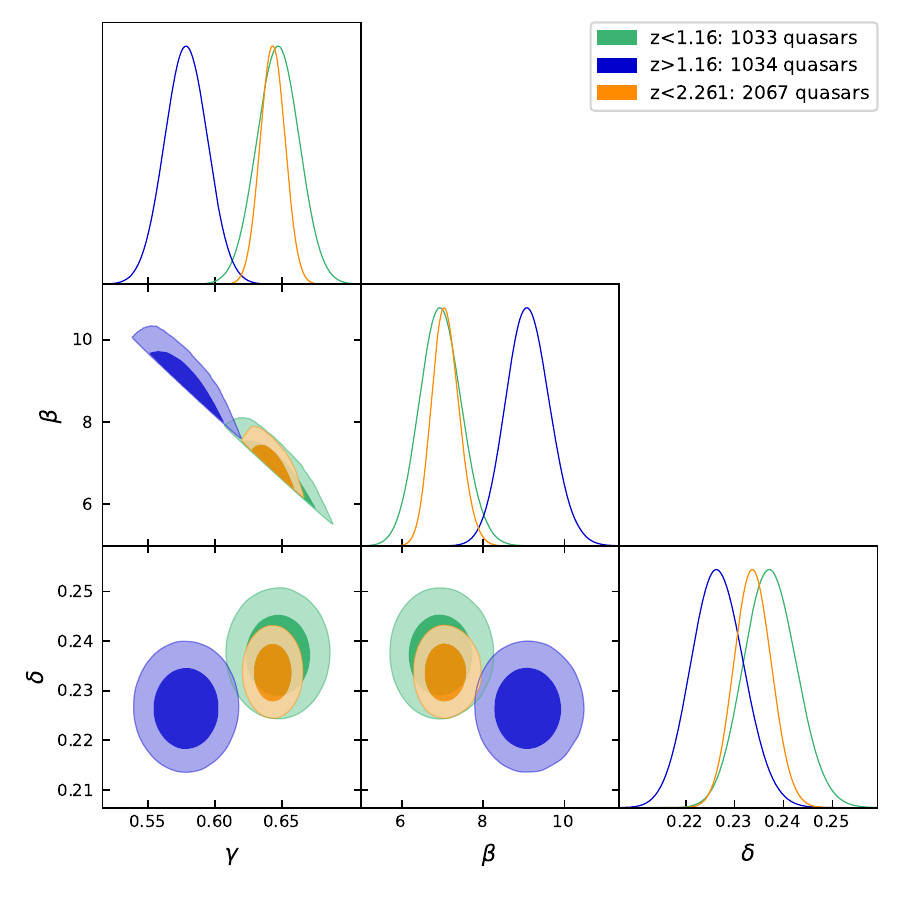}
\caption{Model-independent calibration results for the quasar parameters. GP reconstructions of $D_{\rm{L}}H_0$ from the Pantheon+ SNe Ia compilation were used.
The contours represent the 1$\sigma$, 2$\sigma$ uncertainties for $\gamma , \beta,$ and $\delta$.
Marginal 1-D distributions for each parameter are shown along the diagonal of the corner plot.
}
\label{fig:res_gbd}
\end{figure}

\begin{figure*}
\centering
\includegraphics[width=0.65\textwidth]{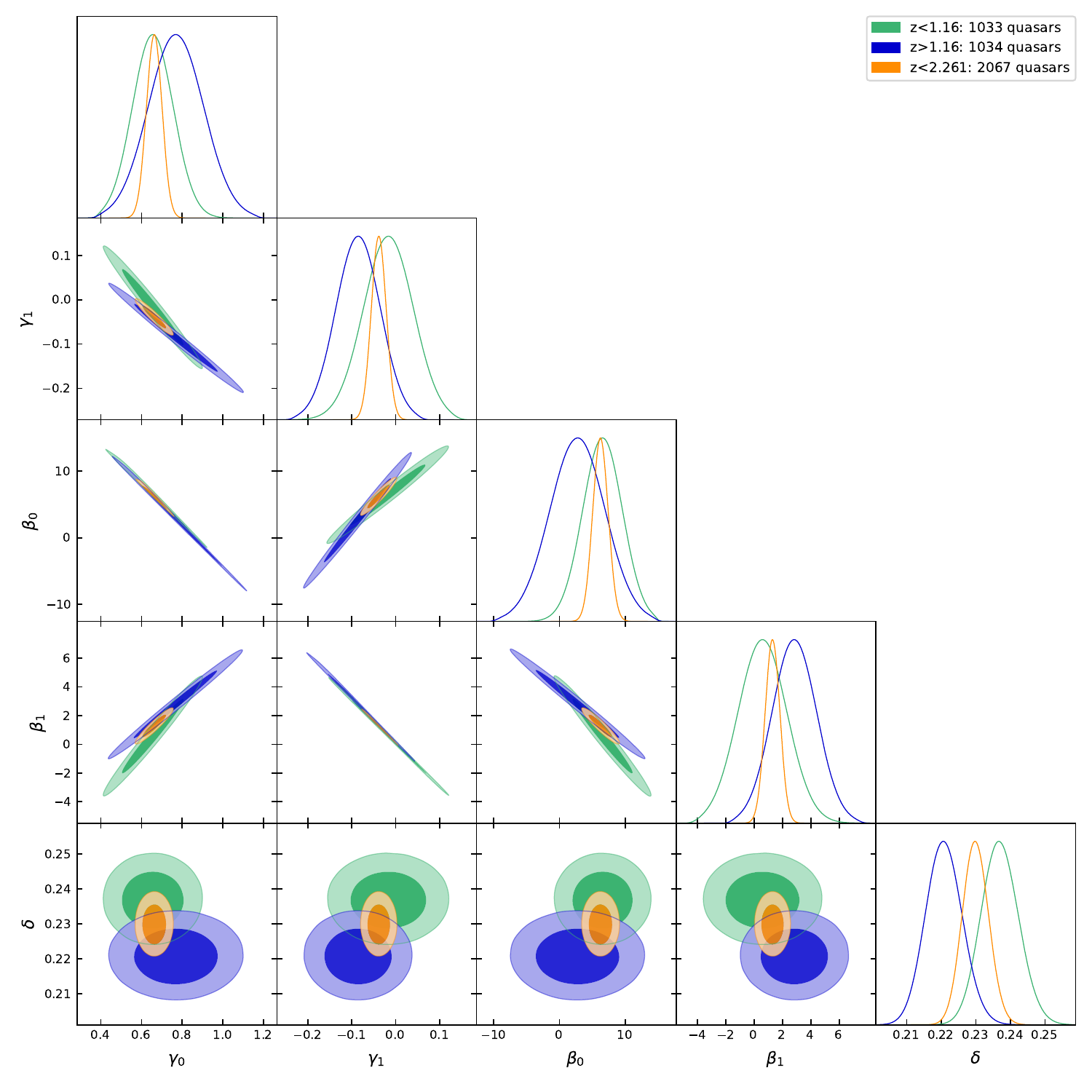}
\caption{
Same as Fig.~\ref{fig:res_gbd} but for the case where the quasar parameters evolve as $\gamma (z)\,=\, \gamma_0+\gamma_1\times (1+z)$ and $\beta (z)\,=\, \beta_0+\beta_1\times (1+z)$.
}
\label{fig:res_gzbzd}
\end{figure*}

\begin{figure}
\centering
\includegraphics[width=0.47\textwidth]{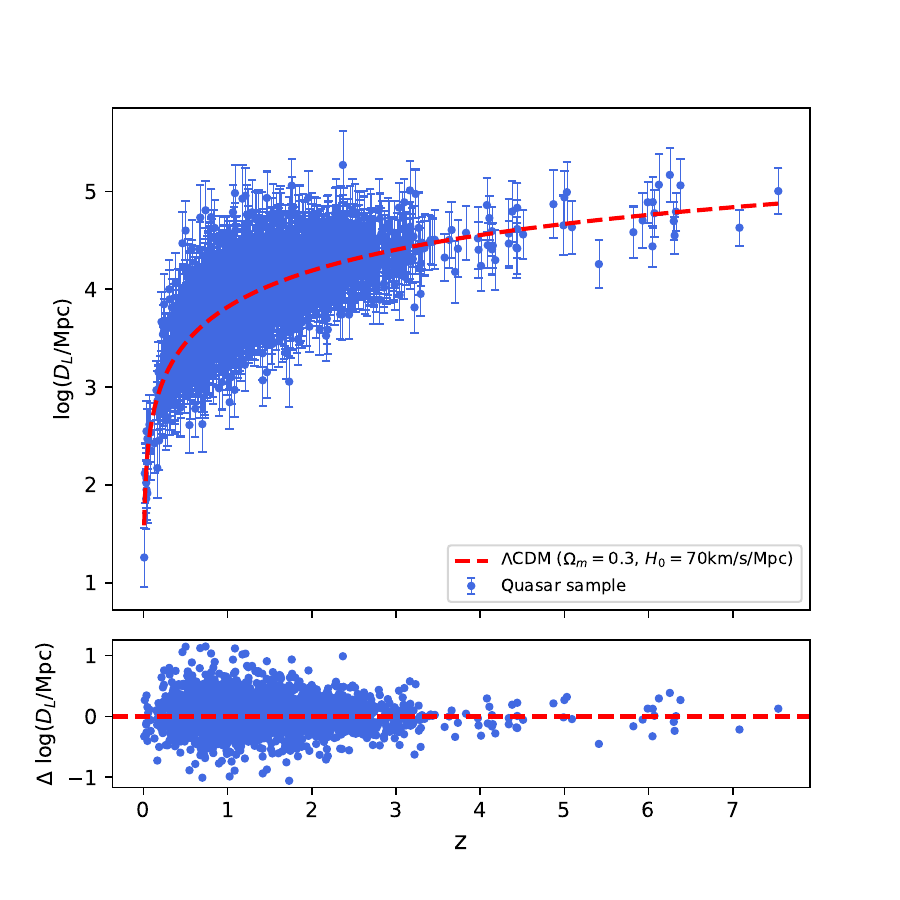}
\caption{The constructed Hubble diagram of the calibrated quasar sample with median values of the quasar parameters.
The upper plot shows the $\log (D_{\rm{L}})$-redshift relation for the 2421 calibrated quasars and the red dashed line shows $\log (D_{\rm{L}})$ from the flat $\LCDM$ model with $\Omega_{\text{m}} = 0.3$ and $H_0 = 70$ km/s/Mpc. The lower plot shows the residuals of $\log (D_{\rm{L}})$ obtained with best-fit quasar sample obtained assuming a flat $\LCDM$ model with $\Omega_{\text{m}} = 0.3$ and $H_0 = 70$ km/s/Mpc.}
\label{fig:res_logDL}
\end{figure}

In this section, we consider two different quasar relations to test the discrepancy between the high-redshift quasar sub-sample and the low-redshift quasar sub-sample.

\subsection{Non-evolution of quasar relation}
Based on the method described in Section~\ref{sec:cali}, we obtain the calibration results with the three quasar sub-samples. The posteriors for the quasar parameters are shown in Figure~\ref{fig:res_gbd}. The green contours (lines) denote the results from the low-redshift sub-sample and the blue contours (lines) denote the results from the high-redshift sub-sample while the orange contours (lines) show the results from the 2067 quasar samples whose redshift range overlap with the redshift range of SNe Ia. As we can see from Figure~\ref{fig:res_gbd}, the constraints from the 2067 quasar sample are consistent with the results from the low-redshift sub-sample while both of them are in tension with the results from the high-redshift quasar sub-sample.  {We can also see a diagonally cut in the $\gamma$ and $\beta$ plane for both the high-z and low-z subsamples, which is mainly attributed to the prior distribution selected for $H_0$. This result suggests that the parameters $\gamma$, $\beta$ and $H_0$ are highly correlated.}

The numerical results including the median values of quasar parameters and the 1$\sigma$ uncertainties are shown in Table~\ref{tab:res_gbd_med}. There is about a 4$\sigma$ deviation for the parameters $\gamma$ and $\beta$ between the high-redshift sub-sample and the low-redshift sub-sample. This deviation indicates that the UV and X-ray luminosity relationship for quasars may evolve with redshift. 
On the other hand, the calibration results for $\gamma$ from the low-redshift sub-sample are consistent with previous work by \cite{Li:2021onq}, which gave $\gamma = 0.649\pm0.007$. However, The calibration results for $\gamma$ from the high-redshift sample indicate $4.4\sigma$ deviation from the previous work by \cite{Li:2021onq}.
We also show the best-fit values of quasar parameters and the $\chi^2$ values in Table~\ref{tab:res_gbd_bf}.

 All the results confirm that there is a tension between the quasar calibration parameters fit at low and high redshifts. Therefore, we should allow for the quasar calibration parameters to evolve with redshift in order to accurately use quasars as standard candles.

\subsection{Redshift-evolution of quasar relation}

In this section, we consider the case for the UV and X-ray luminosity relationship to evolve with redshift following $\gamma = \gamma_0+\gamma_1(1+z)$ and $\beta=\beta_0+\beta_1(1+z)$ in equation~(\ref{eq:logL}), equation~(\ref{eq:logFx1}) and equation~(\ref{eq:logFx}). The calibration results for the quasar parameters are shown in Figure~\ref{fig:res_gzbzd} and the numerical results including the median values of quasar parameters and the $1\sigma$ uncertainties are shown in Table~\ref{tab:res_cali_gammaz1}. We also show the best-fit results for quasar parameters and the accordingly $\chi^2$ in Table~\ref{tab:res_cali_gammaz1bf}. Note that the $\chi^2$ values listed in these tables include the $\sum_i \ln{s_i}$ penalty term as in equation~(\ref{eq:chi2}) and so have negative values.

As we can see from the Figure~\ref{fig:res_gzbzd}, the results for the quasar calibration parameters from the low-redshift sub-sample and the high-redshift sub-sample are consistent at the $2\sigma$ level, which indicates that the parametric form for $\gamma$ and $\beta$ works well in
alleviating the tension between high-redshift and low-redshift quasar sample. However, with the increasing number of quasar calibration parameters to be fit, the uncertainties increase a lot for $\gamma$ and $\beta$. On the other hand, the results from 2067 quasar sample show good consistency with both the low-redshift quasar sub-sample and high-redshift quasar sub-sample, though of course with better precision. However, the uncertainties are still larger than the non-evolution of quasar relation case. {The resolution of the tension observed in quasar parameters may be attributed either to the evolution of the quasar calibration parameters or to host galaxy contamination.
}

{Conversely, the low-redshift sub-sample is consistent with $\gamma_1=0$, $\beta_1=0$, indicating no preference for evolution.  However, the high-redshift sub-sample and the full sample prefer $\gamma_1$ and $\beta_1$ different than 0 at more than 3-$\sigma$. These results indicate that the preference for evolution in the quasar parameters is happening at high redshift and cannot be explained by a $z<0.7$ cut and the corresponding host galaxy contamination.
}

\begin{table*}[]
\caption{The median values of the quasar parameters and their 1$\sigma$ uncertainties as well as the $\chi^2$ from different quasar sub-samples for the case where they do not evolve with redshift. GP reconstructions of $D_{\rm{L}}H_0$ based on the Pantheon+ SNe Ia compilation were used. }\label{tab:res_gbd_med}
\centering
\begin{tabular}{c| c c c c }
\hline
sub-samples & $\gamma$ & $\beta$ & $\delta$ & $\chi^2$ \\
\hline
$z<1.16$: 1033 quasars &$0.647\pm 0.016$ & $6.95\pm 0.52$ &$0.2373\pm 0.0054$  & -1878.88\\
$z>1.16$: 1034 quasars &$0.578\pm 0.016$ & $9.11\pm 0.55$ & $0.2266\pm 0.0054$ & -1929.61\\
\hline
$z<2.26$: 2067 quasars &$0.6431\pm 0.0092$ & $7.08^{+0.31}_{-0.36}$ &$0.2337\pm 0.0038$ & -3760.30\\
\hline
\end{tabular}
\end{table*}

\begin{table}
\caption{The best-fit quasar parameters for the case where they do not evolve with redshift.
GP reconstructions of $D_{\rm{L}}H_0$ based on the Pantheon+ SNe Ia compilation were used.}\label{tab:res_gbd_bf}
\begin{center}
\begin{tabular}{c| c c c c  }
\hline
sub-samples  & $\gamma$ &   $\beta$  & $\delta$ & $\chi^2$\\
\hline
$z<1.16$: 1033 quasars &  $ 0.645 $ & $ 7.56 $   &$ 0.2355$ & -1914.65 \\
$z>1.16$: 1034 quasars &  $0.573 $  & $ 9.34$ &$ 0.2256$ & -1963.84 \\
\hline
$z<2.26$: 2067 quasars &  $0.638 $  & $7.11$ & $0.2331 $ &-3848.34 \\
\hline
\end{tabular}
\end{center}
\end{table}

\begin{table*}
\caption{As Table~\ref{tab:res_gbd_med} but for the case where quasar parameters do evolve with redshift as $\gamma (z)\,=\, \gamma_0+\gamma_1\times (1+z)$ and $\beta (z)\,=\, \beta_0+\beta_1\times (1+z)$.
}\label{tab:res_cali_gammaz1}
\begin{center}
\begin{tabular}{c|  c c c c c c }
\hline
sub-samples  & $\gamma_0$ & $\gamma_1$ & $\beta_0$ &$\beta_1$ & $\delta$ & $\chi^2$\\
\hline
$z<1.16$: 1033 quasars  & $ 0.656\pm 0.099 $ & $-0.017\pm 0.057$ &$ 6.5\pm 3.0 $ &$ 0.6\pm 1.7$ &$0.2370\pm 0.0054$ & -1904.50 \\
$z>1.16$: 1034 quasars  & $0.770\pm 0.130$ &$-0.086\pm 0.050$  &$ 2.7\pm 4.2 $ &$2.8\pm 1.6$ &$0.2208\pm 0.0053 $ & -1991.64 \\
\hline
$z<2.26$: 2067 quasars  & $0.663\pm 0.038 $ & $ -0.039\pm 0.017 $ & $6.2\pm1.2$ & $1.28\pm 0.52$ & $0.2299\pm{0.0038}$ & -3903.08 \\
\hline
\end{tabular}
\end{center}
\end{table*}

\begin{table*}
\caption{As Table~\ref{tab:res_gbd_bf} but for the case where quasar parameters do evolve with redshift as $\gamma (z)\,=\, \gamma_0+\gamma_1\times (1+z)$ and $\beta (z)\,=\, \beta_0+\beta_1\times (1+z)$.
}\label{tab:res_cali_gammaz1bf}
\begin{center}
\begin{tabular}{c| c c c c c c  }
\hline
sub-samples & $\gamma_0$ & $\gamma_1$ & $\beta_0$ &$\beta_1$ & $\delta$ & $\chi^2$\\
\hline
$z<1.16$: 1033 quasars &  0.657 & -0.019 &  6.33 &  0.63 & 0.2357 & -1919.35\\
$z>1.16$: 1034 quasars &  0.757 & -0.082 &  3.13 &  2.71 & 0.2204 & -2006.07\\
\hline
$z<2.26$: 2067 quasars &  0.671 & -0.043 &  6.08 &  1.46 & 0.2295 & -3908.27\\
\hline
\end{tabular}
\end{center}
\end{table*}

\subsection{Hubble diagram of quasars}

Extrapolating the calibrated values of quasar parameters sample and the intrinsic dispersion from the 2067 quasars to the full sample, which include 2421 quasars, we can construct the Hubble diagram of quasars following
\begin{equation}\label{eq:logDL}
\begin{aligned}
{\rm{log }}(D_{\rm{L}}) \,=\,& -\frac{1}{(2\gamma-2)}\times [\gamma \log (F_{\rm{UV}})+ (\gamma-1) \log (4\pi)\\
&+\beta - {\log (F_{\rm{X}})}]
\end{aligned}
\end{equation}
and the errors are obtained from error propagation considering the observed errors of $\log F_{\rm{X}}$ and $\log F_{\rm{UV}}$ as well as the intrinsic dispersion $\delta$. The results are shown in the upper plot of Figure~\ref{fig:res_logDL}. The red dashed line show the ${\rm{log }}(D_{\rm{L}})$ from flat $\LCDM$ model with $\Omega_{\text{m}} = 0.3$ and $H_0 = 70$ km/s/Mpc following
\begin{equation}
    D_{\rm{L}}\,=\,\frac{c}{H_0}(1+z)\int_z^z\frac{dz}{\sqrt{\Omega_{\rm{m}}(1+z)^3+(1-\Omega_{\rm{m}})}}.
\end{equation}
We also calculate the residuals of $\log (D_{\rm{L}})$ obtained with the calibrated quasar sample obtained assuming a flat $\LCDM$ model with $\Omega_{\text{m}} = 0.3$ and $H_0 = 70$ km/s/Mpc. The results are shown in the lower plot of figure~\ref{fig:res_logDL}.

We can see the Hubble diagram is broadly consistent with the flat $\LCDM$ model. However, we will investigate in more detail the cosmological constraints with quasars as standard candles in a companion work.

\section{Conclusions} \label{sec:con}
Quasars are interesting potential standard candles since they would probe redshifts beyond those accessible by SNe Ia. However, there is a significant tension between the quasar calibration parameters if we consider the high-redshift quasars and low-redshift quasars separately. Therefore, we propose a case where the quasar calibration parameters evolve with redshift following $\gamma = \gamma_0+\gamma_1(1+z)$ and $\beta=\beta_0+\beta_1(1+z)$. The quasar sample is then calibrated in a cosmological-model-independent way by using the unanchored luminosity distances $D_{\rm{L}}H_0$ reconstructed from the Pantheon+ SNe Ia data with GP regression.

Our results for the case where the quasar calibration parameters do not evolve with redshift, show that there is about a 4$\sigma$ tension between the quasar parameters $\gamma$ and $\beta$ when fitting the low-redshift quasar sample and the high-redshift quasar sample. In contrast, the case where the quasar calibration parameters are allowed to evolve with redshift, shows that the quasar calibration parameters are consistent at the 2$\sigma$ level.

{Although the uncertainties associated with the quasar calibration parameters increase due to the greater number of these parameters, the internal inconsistencies within the data are resolved. Further, the high-redshift and full samples prefer values of the extended parameters $\gamma_1$, $\beta_1$ different than 0, indicating the restored consistency of the high-redshift and low-redshift sub-samples is due to these evolution parameters and not simply due to increased uncertainties.
While the tension between the high-redshift and low-redshift subsamples are mitigated by employing a conservative cut on quasars at redshift less than 0.7, the high-redshift subsample alone still prefers evolution in the quasar parameters.
These considerations are detailed in the Appendix.
}

In conclusion, quasars could probe the Universe's expansion history at redshifts beyond those probed by SNe Ia, if they can be calibrated accurately.  We have shown that quasars are still sensitive probes of the Universe's expansion history even allowing for more complicated but empirically-driven extensions to the quasars' calibration relationship. This might, in principle, motivate some theoretical studies to discover a more accurate and physically motivated relationship for quasars' UV and X-ray luminosities.  {We hope future surveys with more accurate observations would be able to tell us if there is redshift evolution for quasar UV and X-ray relationship and thus make quasar sample a much more reliable cosmic probe. }


\begin{acknowledgments}
We thank Elisabeta Lusso for valuable suggestions and comments.
This work was supported by National Natural Science Foundation of China (NSFC) No. 12003006 and Science Research Project of Hebei Education Department No. BJK2024134.
This work benefits from the high performance computing clusters at College of Physics, Hebei Normal University. A. S. would like to acknowledge the support by National Research Foundation of Korea NRF2021M3F7A1082056 and the support of the Korea Institute for Advanced Study (KIAS) grant funded by the government of Korea. 
\end{acknowledgments}

\bibliography{sample631}{}
\bibliographystyle{aasjournal}



\begin{figure}[h!]
\centering
\includegraphics[width=0.65\linewidth]{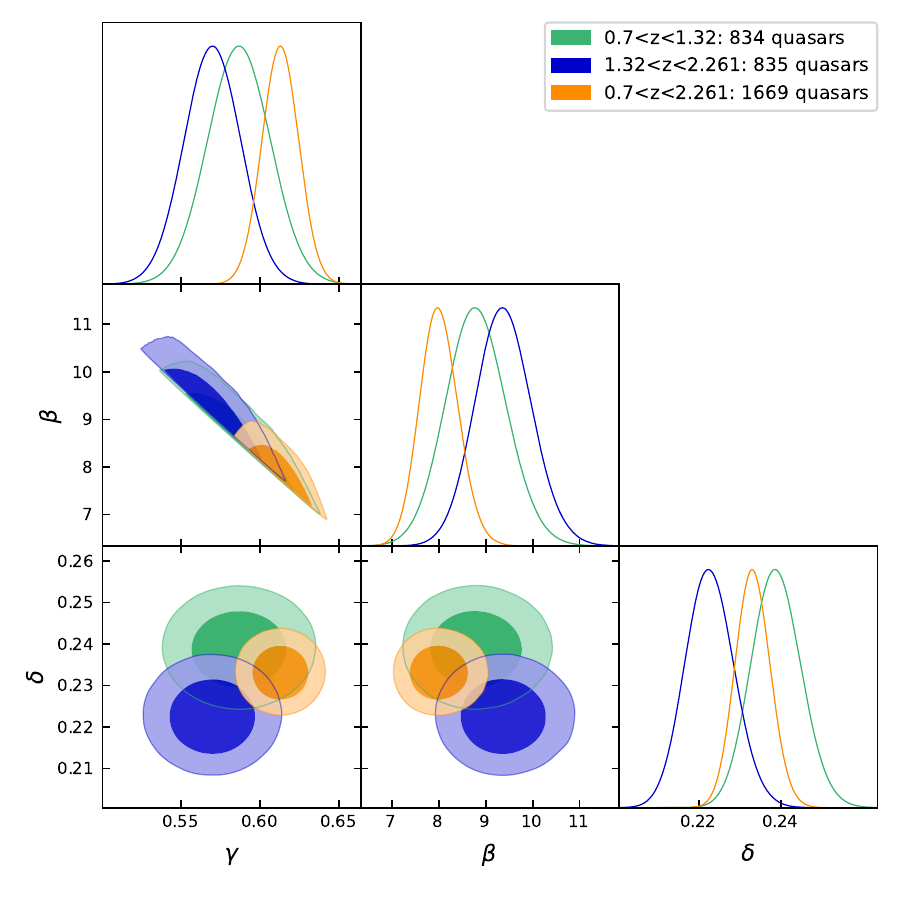}
\caption{\label{fig:NoEvolcon}Calibration results for the quasar parameters with $z<0.7$ quasar excluded for the case where they do not evolve with redshift.}
\end{figure}

\appendix 
After excluding quasars with a redshift below $0.7$, the remaining sample comprises 1669 quasars exhibiting redshift range overlap with SNe Ia. This quasar sample is further stratified into two sub-samples: a low-redshift group consisting of 834 quasars, and a high-redshift group consisting of 835 quasars. Calibration is performed using $D_LH_0$, reconstructed from SNe Ia utilizing GP as outlined in Section~\ref{sec:cali}. The results of this calibration are presented in Figure~\ref{fig:NoEvolcon} and Table~\ref{tab:res_gbd_conser}.

\begin{figure}[h!]
\centering
\includegraphics[width=0.65\linewidth]{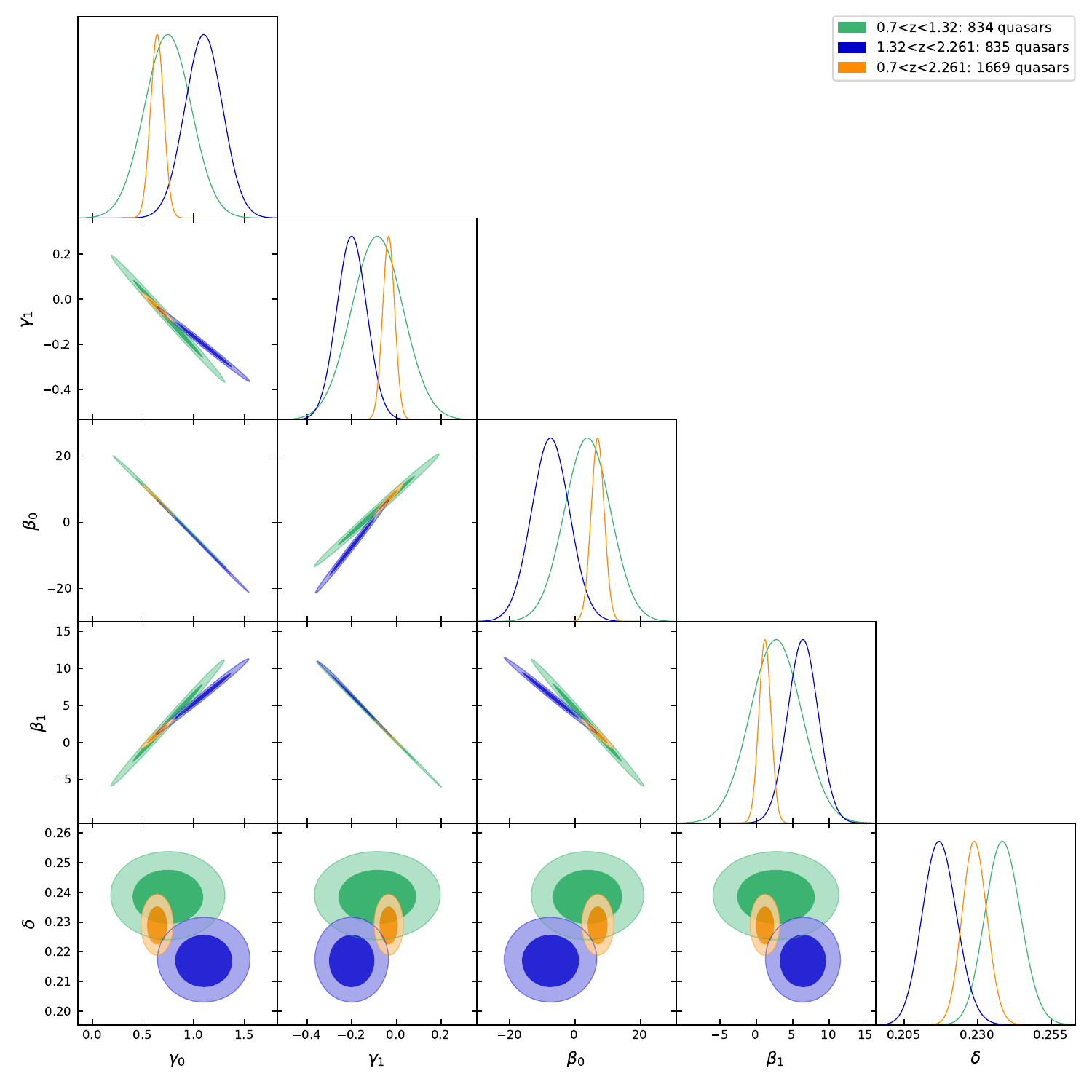}
\caption{\label{fig:Evolconz07} Calibration results for the quasar parameters with $z<0.7$ quasar excluded for the case where they evolve with redshift.}
\end{figure}

\begin{figure}[h!]
\centering
\includegraphics[width=0.30\linewidth]{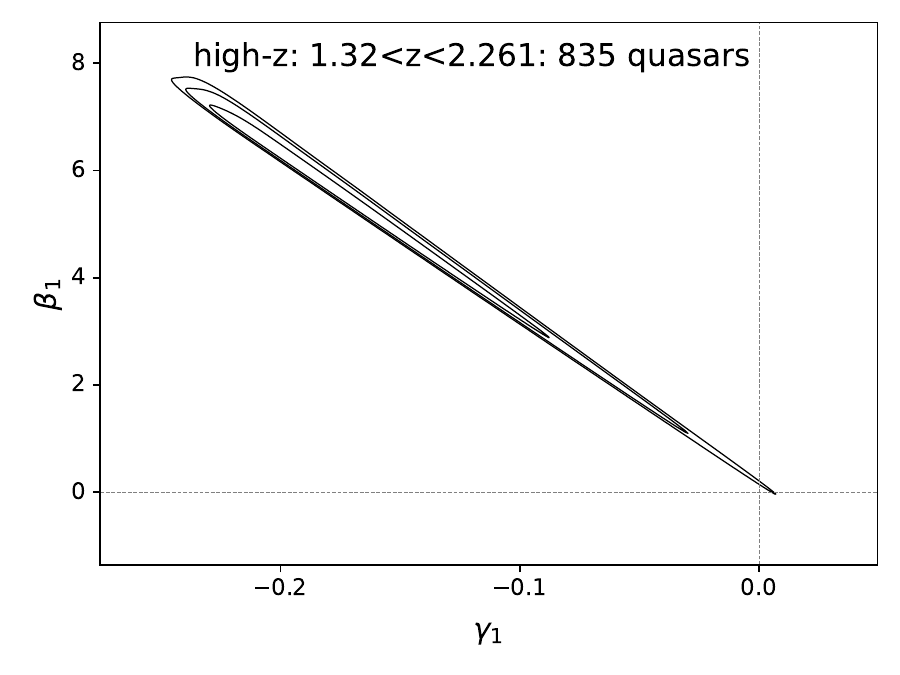}
\includegraphics[width=0.30\linewidth]{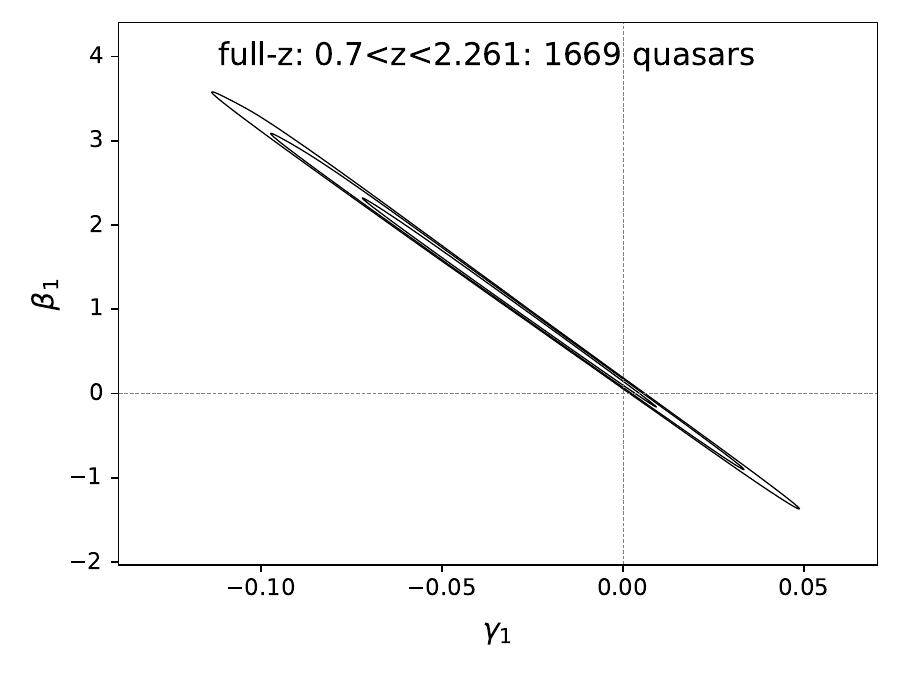}
\includegraphics[width=0.30\linewidth]{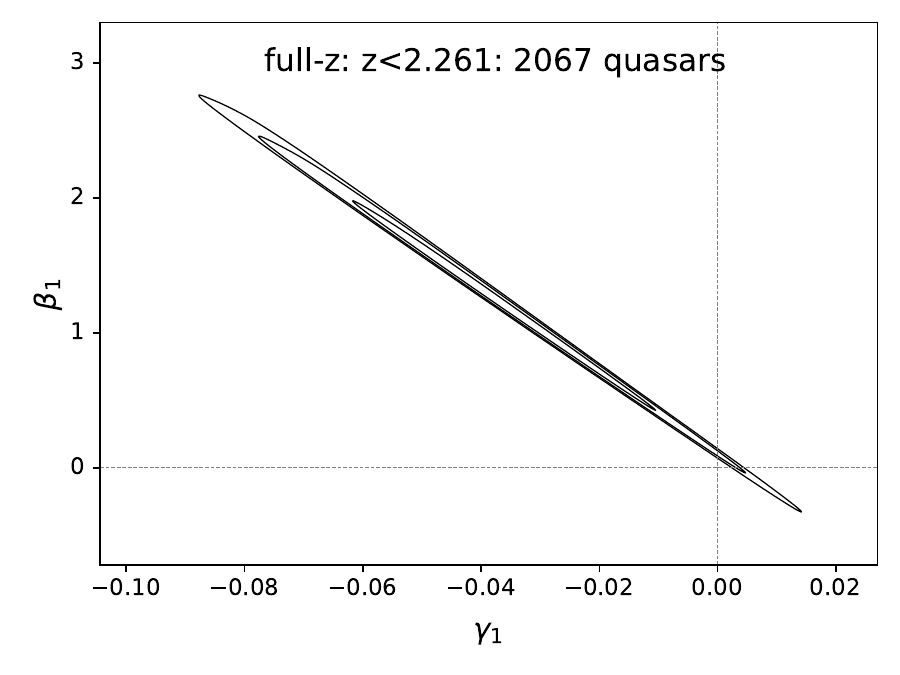}

\caption{\label{fig:Evolcon}The posterior for $\gamma_1-\beta_1$ relation in 2D plots for high-redshift quasar subsamples (left panel), full quasar sample exclude $z<0.7$ quasars (middle panel) and full quasar sample (right panel).}
\end{figure}

The findings indicate that the discrepancies between the low-redshift and high-redshift sub-samples vanish, and the quasar parameters for these sub-samples align within a 1$\sigma$ statistical level of significance, {but this can be due to rigidity of the functional form fitting the data (not being flexible enough). To validate this consistency, we allow then redshift evolution of the quasar parameters. The results are shown in Table~\ref{tab:res_gbdz_07} and Figure~\ref{fig:Evolconz07}. We also show the two-dimension contours for $\gamma_1$ and $\beta_1$ obtained with high-z quasar sub-sample, full quasar sample exclude $z<0.7$ quasars and full quasar sample in Figure~\ref{fig:Evolcon}. We can see that zero slope ($\gamma_1=0,\,\beta_1=0$) are outside the 3$\sigma$ contour for all the cases considered, which indicate that the quasar samples prefers a redshift evolution in the quasar calibration parameters regardless of any z<0.7 contamination.}

\begin{table*}[h!]
\caption{The median values of the quasar parameters and their 1$\sigma$ uncertainties with $z<0.7$ quasar excluded for the case where they do not evolve with redshift. GP reconstructions of $D_{\rm{L}}H_0$ based on the Pantheon+ SNe Ia compilation were used.}\label{tab:res_gbd_conser}
\centering
\begin{tabular}{c| c c c  }
\hline
sub-samples & $\gamma$ & $\beta$ & $\delta$  \\
\hline
$0.7<z<2.26$: 1669 quasars &$0.613\pm 0.012    $&  $8.01^{+0.38}_{-0.43}$& $ 0.2332\pm 0.0043      $\\
\hline
$0.7<z<1.32$: 834 quasars &$0.587\pm 0.020  $ & $8.80\pm 0.65    $ & $0.2389\pm 0.0061$\\
$1.32<z<2.26$: 835 quasars & $0.570\pm 0.018    $& $9.38\pm 0.60    $& $0.2227\pm 0.0060$\\
\hline
\end{tabular}
\end{table*}

\begin{table*}[h!]
\caption{As Table~\ref{tab:res_gbd_conser}, but for the case where quasar parameters do evolve with redshift as $\gamma (z)\,=\, \gamma_0+\gamma_1\times (1+z)$ and $\beta (z)\,=\, \beta_0+\beta_1\times (1+z)$.}\label{tab:res_gbdz_07}
\centering
\begin{tabular}{c| c c c c c  }
\hline
sub-samples &  $\gamma_0$ & $\gamma_1$ & $\beta_0$ &$\beta_1$ & $\delta$  \\
\hline

$0.7<z<2.26$: 1669 quasars&$0.638\pm 0.065  $     &$-0.034\pm 0.027 $   &$6.9\pm 2.0      $ &$1.18\pm 0.83    $ &$0.2290\pm 0.0042$ \\
\hline
$0.7<z<1.32$: 834 quasars  & $0.740\pm 0.230    $ &  $-0.090\pm 0.120   $ &  $3.8\pm 7.0$ &        $2.7\pm 3.5      $ & $0.2387\pm 0.0061$ \\
$1.32<z<2.26$: 835 quasars & $1.100\pm 0.190    $ &   $-0.200\pm 0.067 $ &   $-7.5\pm 5.8     $ & $6.4\pm 2.1      $ &  $0.2172\pm 0.0058$ \\

\hline
\end{tabular}
\end{table*}

\begin{figure}[h!]
\centering
\includegraphics[width=0.65\linewidth]{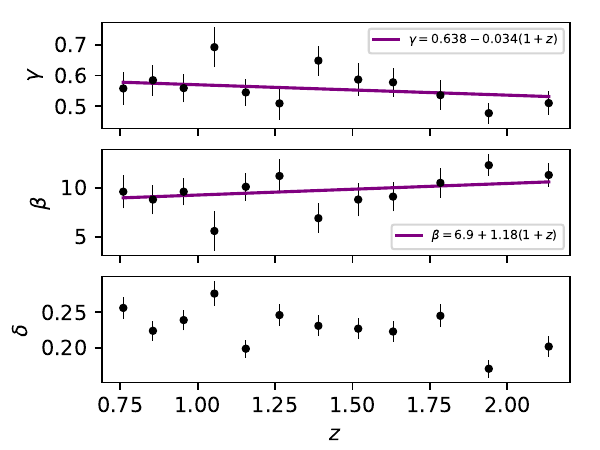}
\caption{The constrain results for quasar parameters with each quasar bins.\label{fig:gbbins}    }
\end{figure}

{{We further stratified the quasar sample with $z>0.7$ into 13 redshift bins. Utilizing the mean values of 1,000 GP reconstructed $D_LH_0$ realizations from SNe Ia, we systematically determined the quasar parameters ($\gamma$, $\beta$) across all bins. The bin-specific sample characteristics and calibration outcomes are comprehensively presented in Table~\ref{tab:zbins} and Figure~\ref{fig:gbbins},  with the latter demonstrating a distinct negative correlation for $\gamma$ and positive correlation for $\beta$ that aligns with the tabulated results.}}

\begin{table*}[h!]
\caption{The detail of the quasar bins including the number of the quasar, the redshift range and the mean redshift for each quasar bins. The last two columns give the constrain results from each quasar bins. }\label{tab:zbins}
\centering
\begin{tabular}{ c c c c c  }
\hline
number of quasars & $ z$  & $ z_{\rm{mean}}$ & $\gamma$& $\beta$   \\
\hline
138 & $0.712 <z< 0.805 $ & 0.7596 &  $0.558\pm0.054$ & $9.6 \pm1.7$ \\   
138 & $0.805 <z< 0.906 $ & 0.8554 &  $0.585\pm0.051$ & $8.8 \pm1.5$ \\   
138 & $0.9066<z< 1.0005$ & 0.9547 &  $0.559\pm0.047$ & $9.6 \pm1.4$ \\
138 & $1.0021<z< 1.1099$ & 1.0539 &  $0.693\pm0.065$ & $5.6 \pm2.0$ \\
138 & $1.1105<z< 1.204 $ & 1.1555 &  $0.545\pm0.044$ & $10.1\pm1.4$ \\  
138 & $1.2046<z< 1.3311$ & 1.2641 &  $0.509\pm0.054$ & $11.2\pm1.7$ \\
138 & $1.332 <z< 1.4569$ & 1.3898 &  $0.649\pm0.049$ & $6.9 \pm1.5$ \\
138 & $1.4592<z< 1.574 $ & 1.5183 &  $0.587\pm0.053$ & $8.8 \pm1.7$ \\
138 & $1.5751<z< 1.703 $ & 1.6309 &  $0.578\pm0.048$ & $9.1 \pm1.5$ \\
138 & $1.703 <z< 1.869 $ & 1.7838 &  $0.536\pm0.048$ & $10.5\pm1.5$ \\
138 & $1.87  <z< 2.0318$ & 1.9402 &  $0.477\pm0.034$ & $12.3\pm1.1$ \\
135 & $2.0336<z< 2.261 $ & 2.1337 &  $0.510\pm0.038$ & $11.3\pm1.2$  \\
\hline
\end{tabular}
\end{table*}

\end{document}